\documentclass[useAMS,usenatbib]{mn2e}
\usepackage{graphicx}
\usepackage{amssymb}
\usepackage{aas_macros}

\def\kms{\;{\rm km\,s}^{-1}}

\def\nbody{$N$-body\,}
\title[Core Radii of star clusters]
  {Core radius evolution of star clusters}
\author[M. I. Wilkinson et al.]
  {M.~I.~Wilkinson$^1$, J.~R.~Hurley$^2$, A.~D.~Mackey$^1$, G.~Gilmore$^1$,
C.~A.~Tout$^1$\\
  $^1$Institute of Astronomy, Madingley Rd, Cambridge CB3 0HA\\
  $^2$Department of Astrophysics,American Museum of Natural History,Central Park West at 79th Street,
New York, NY 10024, USA}
\date{Released 2003 Xxxxx XX}

\pagerange{\pageref{firstpage}--\pageref{lastpage}} \pubyear{2003}

\def\LaTeX{L\kern-.36em\raise.3ex\hbox{a}\kern-.15em
    T\kern-.1667em\lower.7ex\hbox{E}\kern-.125emX}
\begin{document}

\label{firstpage}

\maketitle

\begin{abstract} 
We use \nbody simulations of star clusters to investigate the possible
dynamical origins of the observed spread in core radius among
intermediate-age and old star clusters in the Large Magellanic Cloud
(LMC). Two effects are considered, a time-varying external tidal field
and variations in primordial hard binary fraction. Simulations of
clusters orbiting a point-mass galaxy show similar core radius
evolution for clusters on both circular and elliptical orbits and we
therefore conclude that the tidal field of the LMC has not yet
significantly influenced the evolution of the intermediate-age
clusters. The presence of large numbers of hard primordial binaries in
a cluster leads to core radius expansion; however, the magnitude of
the effect is insufficient to explain the observations. Further, the
range of binary fractions required to produce significant core radius
growth is inconsistent with the observational evidence that all the
LMC clusters have similar stellar luminosity functions.
\end{abstract}


\section{Introduction}
\label{sec:introduction}
Star clusters are vital laboratories for testing theories of stellar
dynamics and stellar evolution~\citep[e.g.][]{mh97}. The globular
clusters orbiting the Milky Way constitute a sample of about $150$
evolved stellar systems with ages $\gtrsim 10^{10}$ years. These
provide useful constraints on the long term evolution of clusters but
information about the paths which led to these end states can only be
inferred via comparisons with the end points of numerical cluster
simulations. In contrast, the system of rich star clusters in the
Large Magellanic Cloud (LMC) includes clusters of comparable mass to
Galactic globular clusters but which span the entire age range from
less than $5\times 10^{6}$ years (e.g. R 136) to about $10^{10}$ years
(e.g. NGC 2019). This allows us to observe massive star clusters at
various epochs in their evolution and makes it possible to identify
evolutionary trends, under the assumption that the external processes
affecting cluster evolution have not changed significantly over the
lifetime of the LMC. With this aim in mind, HST project
GO7307~\citep[e.g.][]{be99} obtained deep images of four pairs of LMC
clusters. Within each selected pair the clusters are approximately
coeval, have similar masses and metallicities and lie at comparable
radii from the LMC centre but differ significantly in other properties
such as core radius. The pairs range in age from about $10^7$ years up
to $10^{10}$ years. The youngest clusters (NGC 1805 and NGC 1818) are
old enough to have expelled all their residual gas and so are pure
\nbody systems.

\begin{figure}
\includegraphics[width=84mm]{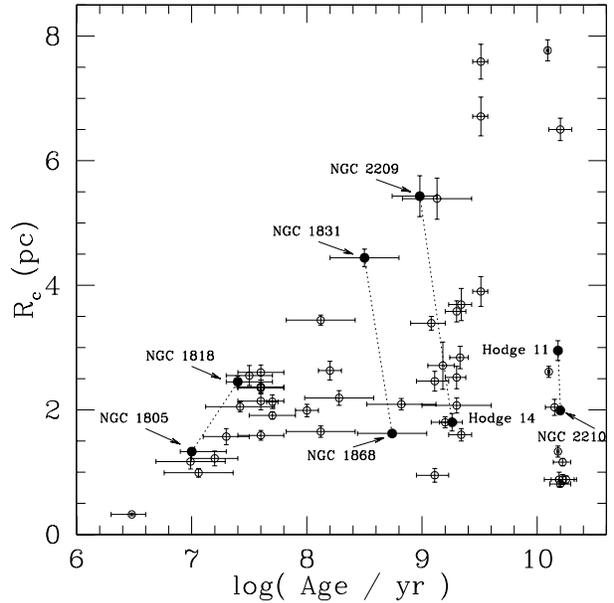} 
\caption{Observed core radii $R_{\rm c}$ of LMC star clusters as a function
of age. The clusters in the GO7307 HST sample are identified by name, and
matched pairs are joined by dotted lines. (Following Fig.~1 of
\protect\cite{deg02a}, data taken from \protect\cite{mg03}).}
\label{fig:lmccores}
\end{figure}

Fig.~\ref{fig:lmccores} shows a plot of the observed core radii of
star clusters in the LMC as a function of their ages, with the
clusters in the GO7307 HST sample identified by name. There is a clear
trend that the spread in core radius is an increasing function of
cluster age. This trend was first identified by~\cite{efl89} from
ground-based images and was recently confirmed by~\cite{mg03} using
HST archive data.~\cite{mg03} calculated core radii for $53$ LMC
clusters in a uniform manner with the aim of obtaining an unbiased
data set. They discuss in some detail the uncertainties in the data
and conclude that selection effects are not responsible for the
observed trend. They suggest, as~\cite{efl89} had earlier argued, that
the increased spread in core radius is an evolutionary phenomenon: all
clusters formed with relatively small core radii ($\sim 2-3$ pc) and
subsequently some clusters experienced core expansion while others did
not.

\cite{efl89} discuss the possibility that inter-cluster variations in
the slope of the initial mass function (IMF) could have produced the
observed spread in core radius. They point out that mass loss from the
increased numbers of massive stars found in clusters with IMFs flatter
than the standard Salpeter IMF could lead to core expansion of the
observed magnitude on a timescale of about $10^{8.5}$
years. As~\cite{el91} notes, however, the IMF slope required to
explain the largest core radii would lead to disruption of the
clusters on a timescale of a few $\times 10^{7}$ years. More directly,
based on a detailed observational analysis of the clusters in the
GO7307 HST sample,~\cite{deg02a} have shown that the slopes of their
luminosity functions are similar to each other within the
observational uncertainties, strongly suggesting that their IMFs were
very similar down to stellar masses of about $0.8-1.0$M$_\odot$. The
absence of inter-cluster population variations implies that stellar
evolutionary mass loss cannot be invoked to explain the observations.

As a star cluster evolves, encounters lead to a redistribution of
energy among the stars. Heavier stars and binaries acquire lower
velocities, on average, and sink towards the cluster core. This
process of mass segregation could enhance the heating effect of a
population of primordial binaries because the binaries would tend to
fall to the centre where encounters occur more
frequently. Observations of young clusters~\citep[e.g.][]{hillen97}
and numerical simulations of star formation~\citep[e.g.][]{bbcp01}
show that so-called primordial mass segregation can also result from
the cluster formation process.~\cite{deg02a} have shown that there is
strong evidence for mass segregation in all the young and
intermediate-age clusters in the GO7307 sample. Further, the amount of
mass segregation appears to be similar in all clusters and cannot
therefore account for differences in the cluster core radii.

The aim of this paper is to investigate, by means of \nbody
simulations, whether dynamical processes could give rise to the
observed core radius--age relation. In this context, the pair of
clusters NGC 1868 and NGC 1831 are of particular interest.  These
clusters have very similar ages and metallicities and their masses
differ by at most a factor of two~\citep[see Table~1
of][]{deg02a}. However, as Fig.~\ref{fig:lmccores} shows, the core
radius of NGC 1831 is almost three times larger than that of NGC
1868. Given that the cluster ages are only around $500\,$Myr any
process which has affected the core radii must be capable of producing
significant evolution even on this relatively short time scale. In
this paper, we consider the dynamical effects of (1) variations in the
external tidal field experienced by the clusters and (2) differences
in the initial numbers of dynamically active binary stars in the
clusters. Given that the crossing times at the half-mass radii of
these clusters are about $10^{7}$ yrs and the times to orbit the LMC
are about the same order of magnitude, it is reasonable to suppose
that either of these processes could have begun to produce observable
effects on cluster structure on the relevant time scale.

During the past decade, numerical simulations of star clusters have
become increasingly realistic owing to dramatic increases in
computational speeds and significant improvements in programming
algorithms~\citep[e.g.][]{aa99}. The \nbody code used in this paper
incorporates the stellar evolution models of~\cite{hpt00,htp02}. This makes
it possible to analyse the output from the simulations in an identical
manner to that used for observational data, thus allowing meaningful
confrontation of models with observations. This is particularly
important when discussing cluster properties such as core radius
because the relation between the theoretical and observational
definitions of this quantity depend on the conversion between the mass
function and luminosity function of the cluster stars. For a real
cluster this conversion is non-trivial because it depends on, among
other quantities, the cluster metallicity and is further complicated
by the possible presence of mass segregation~\citep[see
e.g.][]{deg02b}. In this paper, all comparisons between models and
observations are made in terms of directly observable properties.

In the interests of clarity, we include here a note on the definition
of the core radius. As~\cite{ch86} discuss, theoreticians, numerical
simulators and observers assign different meanings to the term ``core
radius'', reflecting the quantity of greatest relevance to their
purpose. In theoretical work, the core radius $r_{\rm c}$ refers to
the natural length scale of the system under study, for example the
scale radius of the widely used~\cite{king66} profile. The operational
definition of the core radius commonly used in discussions of
\nbody$\,$ simulations is that of the density radius $r_{\rm d}$. This
is the density-weighted average of the distance of each star from the
density centre~\citep{aa01}, where the local density at each star is
estimated from the mass within a sphere containing the star's six
nearest neighbours and the density centre is defined as the
density-weighted mean position of the stars. Observationally, since
the available information is either in the form of cluster surface
brightness profiles or luminosities of individual stars instead of
mass densities and stellar masses, the core radius $R_{\rm c}$
corresponds to the radius at which the projected surface brightness
falls to half its central value. In the case of a cluster of
equal-mass stars in equilibrium, there is generally a well-defined
relation between $r_{\rm c}$ and $R_{\rm c}$. For example, for
a~\cite{plum11} model it is straightforward to show that $R_{\rm c} =
0.6435 r_{\rm c}$. It is important to note, however, that during the
evolution of the cluster this relation is not static and that there is
no general relation between $r_{\rm d}$ and either $r_{\rm c}$ or
$R_{\rm c}$. In this paper, we use the term ``core radius''
exclusively to refer to the observational quantity $R_{\rm c}$. In
Section~\ref{sec:tides_discussion} we will compare the evolution of
$r_{\rm d}$ and $R_{\rm c}$, in order to emphasize the care which must
be taken in comparing observations with simulations.

In Section~\ref{sec:nbody} we describe briefly the \nbody code used in
the simulations, as well as the pipeline used to calculate
observational properties of the simulated
clusters. Section~\ref{sec:tides} considers the effects of time
variation of an external tidal field on the evolution of the cluster
core radius while Section~\ref{sec:binfrac} deals with clusters moving
in a steady tidal field but with different initial binary
fractions. In Section~\ref{sec:bintide} we investigate the combined
effect of both mechanisms by studying binary-rich clusters in a
time-varying tidal field and the results of these simulations are
discussed in the context of the LMC clusters in
Section~\ref{sec:bintidedisc}. Finally in
Section~\ref{sec:conclusions} we summarise our conclusions.

\section{$N$-body simulations}
\label{sec:nbody}
\begin{table}
\begin{minipage}{84mm}
  \caption{Summary of \nbody runs performed. $N_0$, $M_0$ and $f_{{\rm
b},0}$ are the initial number of stars, total stellar mass and
primordial fraction of hard binaries, respectively. The clusters were
all simulated in the tidal field of a point mass representing the LMC,
and each simulation ran until time $T_{\rm max}$. }
\label{tab:runs} \begin{tabular}{lcccccc} \hline Name
& $N_0$ & $M_0$ (M$_\odot$) & $f_{{\rm b},0}$ & Orbit & $T_{\rm
max}$\\ \hline Circ & $5000$ & $2390$ & 0 & Circular & $1.0$ Gyr\\ Ecc
& $5000$ & $2410$ & 0 & Elliptical & $1.0$ Gyr\\ Binary & $4500$ &
$2310$ & $0.5$ & Circular & $1.7$ Gyr\\ Binary2 & $4500$ & $2390$ &
$0.25$ & Circular & $1.6$ Gyr\\ Binary3 & $4500$ & $2180$ & $0.1$ &
Circular & $1.0$ Gyr\\ BinaryEcc & $4500$ & $2340$ & $0.5$ &
Elliptical & $1.6$ Gyr\\ \hline \end{tabular}
\end{minipage}
\end{table}
\nbody simulations are a very valuable tool in the study of star
clusters because they allow many different processes to be studied
with a minimum of simplifying assumptions. The simulations discussed
below were performed using the NBODY4 code~\citep[e.g.][]{aa99}
running on the HARP-3 and GRAPE-6 special purpose
computers~\citep{mak97} at the Institute of Astronomy, Cambridge and
on a GRAPE-6 board at the American Museum of Natural History.  The
NBODY4 code uses the 4th order Hermite integration scheme to follow
the orbits of the cluster stars and utilises state-of-the-art
regularisation schemes~\citep{ma93,ma98} for the accurate and
efficient treatment of the internal motion of binary and higher-order
subsystems. It also incorporates a full set of stellar evolution
models, in the form of parameterized functions, which are used to
follow the evolution of single stars and binaries~\citep{htap01}.  Of
particular importance to the present study is the treatment of mass
loss. Any gas which is lost from an evolving star and is not accreted
on to a binary companion is assumed to leave the cluster
instantaneously, thereby reducing the depth of the cluster potential
well. A realistic treatment of mass loss is therefore essential to
allow the study of the evolution of such cluster properties as the
core radius.

Within the \nbody code the equations of motion are integrated in
scaled units such that $G=1$ and the total mass and virial radius of
the cluster are also set to unity~\citep{hm86}. For a star cluster in
virial equilibrium the initial energy in these units is $-1/4$ and the
crossing time is $2\sqrt{2}$. Given the total mass of the cluster in
solar masses and an appropriately chosen length scale (which
determines the conversion from \nbody length units to physical units)
it is straightforward to obtain the conversion factors for time and
velocity from \nbody units to Myr and $\kms$, respectively.

In order to facilitate direct comparison of the results of our
simulations with the observed data, we determine the core radius for
each model cluster using the same processing pipeline as that used
by~\cite{mg03}. In summary, the stellar data from the model clusters
(in the form of position and luminosity information for each member
star) are binned radially to obtain a projected surface brightness
profile. The three-parameter model~\citep{eff87}
\begin{equation}
\mu (r) = \mu_0 \left( 1 + \frac{r^2}{a^2} \right)^{-\frac{\gamma}{2}}
\end{equation}
is then fitted to the cluster profile. Here, $\mu_0$ is the central
surface brightness, $\gamma$ determines the fall-off of the surface
brightness profile at large radii and $R_{\rm c}$ is related to the
length scale $a$ via $R_{\rm c} = a(2^{2/\gamma}-1)^{1/2}$. Full
details of the fitting procedure used and discussion of the associated
uncertainties are presented in~\cite{mg03}. In order to reduce the
numerical noise due to the small number of stars in our simulated
clusters we observe each cluster along three perpendicular lines of
sight and obtain the median core radius from these three
estimates. This helps to reduce the impact of individual bright stars
which might otherwise skew the core radius estimation. Further, each
\nbody run is performed four times using different random realisations
of the initial cluster model and the results are averaged in order to
ensure the robustness of any observed trends.

Initially, all the model clusters are generated according to
a~\cite{plum11} density law 
\begin{equation}
\label{eqn:plummer}
\rho(r) = \frac{\rho_0 r_{\rm P}^5}{(r_{\rm P}^2 + r^2)^{5/2}},
\end{equation}
following the prescription of~\cite{ahw74}. Here $r_{\rm P}$ is the
Plummer length scale and $\rho_0$ is the central mass
density. Observations of the young clusters in the LMC show that they
do not appear to be tidally limited~\citep{mg03}. In this case, the
Plummer profile is a reasonable model to use for the initial density
profile given that the fall-off in its outer regions lies between that
of a tidally truncated King model and the observed profiles of the
young LMC clusters.

The initial distribution of stellar masses is drawn from the IMF
of~\cite{ktg93} which is a three-part power law fit to stellar data in
the solar neighbourhood. We take the lower and upper mass limits to be
$0.1$M$_\odot$ and $50$M$_\odot$, respectively. We assume solar
metallicity for the stars in our clusters. This is an appropriate
choice for the younger clusters -- for example,~\cite{jbg01} have
shown that NGC 1805 and NGC 1818, whose ages are about $10$ and
$25\,$Myr respectively, are well matched by isochrones of
approximately solar metallicity. Since the pairs of clusters in the
GO7307 HST sample were chosen to have similar metallicity, we know a
priori that the observed range of cluster core radii cannot be due to
metallicity variations. We have therefore chosen to perform all our
simulations at solar metallicity, and postpone discussion of the
effects of metallicity variations on star cluster evolution to future
work.

Table~\ref{tab:runs} summarises the simulations which we have
performed. The clusters in these runs initially contain $4500$ or
$5000$ stars and have a mean mass of about $2300$ M$_\odot$. Clearly,
the LMC clusters are more massive and contain greater numbers of stars
than our model clusters and therefore direct comparison of the results
of our simulations with the observed data must be approached with
caution. However, the effects of increasing $N$ in our simulations can
be estimated by considering the various time scales in the problems we
study. Detailed discussion of these scaling issues is presented in
Sections~\ref{sec:tides_discussion} and~\ref{sec:bintidedisc} where we
argue that none of our conclusions is strongly dependent on the small
size of our simulated clusters.

\section{Time-varying tidal fields}
\label{sec:tides}
\subsection{Model Initialisation}
\label{sec:tides_models}
We first consider the effects of the external tidal field of the LMC
on the evolution of the cluster core radii. Theoretical arguments and
numerical simulations strongly suggest that tides have played a
significant role in the evolution of star clusters and cluster systems
in galaxies such as the Milky
Way~\citep[e.g.][]{gno97,vh97,ve98,baum99,fz01} . In addition,
observations of tidal tails around at least twenty Milky Way globular
clusters~\citep[e.g.][]{lmc00,od01} provide compelling evidence of the
ongoing disruption of star clusters by the Milky Way. The
gravitational field of the LMC (mass $\approx 10^{10}$M$_\odot$) is
much weaker than that of a galaxy like the Milky Way (mass $\approx
10^{12}$M$_\odot$). However, as~\cite{mg03} show (their Fig.~15),
there is a deficit of low-mass clusters older than $10^{10}$ years. If
we assume that the cluster population originally contained a spectrum
of cluster masses similar to that of the present-day young clusters,
their absence at late times strongly suggests that tides have been at
work on the cluster system in exactly the manner expected from the
models of~\cite{gno97}. Given that the GO7307 HST sample of 8 clusters
was selected on the basis that they lie at similar distances from the
LMC centre but display a wide range of core radii, the steady tidal
field experienced by a cluster on a roughly circular orbit cannot be
the origin of the observed trends. Further, given the intrinsic
weakness of the LMC field at a radius of $4\,$kpc, where our clusters
lie, one might not expect the tidal field to play a significant role
in cluster evolution. We therefore consider clusters on non-circular
orbits. Such clusters experience tidal shocks during their orbits and
we investigate whether these impart sufficient energy to the cluster
to produce differences in the evolution of $R_{\rm c}$.~\cite{el00}
discusses the possible role of disk shocking in the internal evolution
of the intermediate-age LMC clusters, but questions whether any of the
clusters is on an appropriate disk-crossing orbit for this to be an
important effect. In this section we consider instead the tidal
shocking produced by a point mass galaxy.

In order to investigate the role of the LMC tidal field in the
evolution of its cluster system, it is necessary to incorporate an
appropriate external gravitational field into the simulations. We
model the LMC potential as that of a point mass of $9\times
10^{9}$M$_\odot$. This is a rather drastic oversimplification of the
actual situation. We note, however, that this model exaggerates the
effect of the tidal field, as any extended distribution of mass will
have a more slowly varying gravitational potential.  Further, in the
radius range $2-8$~kpc, in which our model clusters orbit, the
gradient of the potential is within a factor of two of that of the
detailed LMC mass model of~\cite{vdm02}. Thus our simulations
constitute a plausible upper limit to the possible effects of tides.

In the \nbody simulations, the equations of motion are integrated in
an accelerating but non-rotating reference frame centred on the centre
of mass of the cluster and whose axes are aligned with fixed spatial
directions. This choice of frame simplifies the equations of motion
for a cluster on a non-circular orbit despite having less symmetry
than the more usual rotating frame used for steady tidal fields whose
axes always point towards the attracting centre~\citep{he01}.
Implementation of the time varying tidal field requires some care
because mass loss from evolving stars or the escape of stars from the
cluster causes the centre of mass to acquire a non-zero velocity
relative to the frame of integration. It is therefore necessary to
re-centre the frame of reference to coincide with the rest frame of
the centre of mass following each such event. All internal and
external forces must then be reinitialized to ensure consistency with
the re-centred frame during subsequent integrations.

We have performed simulations for clusters on elliptical orbits with
perigalactic radii $R_{\rm p} \approx 2\,$kpc and apogalactic radii
$R_{\rm a} \approx 8\,$kpc, probing the region in which most of our
sample clusters lie but largely avoiding the complications of the
inner regions of the LMC which may contain a stellar
bar~\citep[e.g.][]{ez00}. We also modelled clusters on circular orbits
at a radius of $6\,$kpc which corresponds to the time-averaged radius of
the elliptical orbits. This allows us to isolate the effect of the
time variation of the tidal field because both sets of clusters
experience the same mean tidal field. 

We calculate the initial perigalactic tidal radius $r_{\rm t,p}$ in
the approximation that the cluster is on a circular orbit at a radius
$R_{\rm p}$. In this case, $r_{\rm t,p}$ is determined from the
cluster mass $M$, the mass of the point mass galaxy $M_{\rm G}$ and
$R_{\rm p}$ via~\citep[e.g.][]{king62}
\begin{equation}
r_{\rm t,p} = R_{\rm p} \left( \frac{M}{3 G M_{\rm G}} \right)^{1/3} .
\end{equation}
The apocentric tidal radius $r_{\rm t,a}$ is determined in a similar
way from $R_{\rm a}$. The unit of length, which determines the
conversion between \nbody length units and physical units, is then
chosen such that the limiting radius of all clusters is initially
equal to $r_{\rm t,p}$. However, at the start of the simulations the
clusters are placed at the apocentre of their orbits about the LMC
which means that they do not fill their tidal radii. If, instead, the
clusters initially extended to $r_{\rm t,a}$ and were not truncated at
$r_{\rm t,p}$ then the removal of stars outside $r_{\rm t,p}$ during
the first few perigalactic passages could mask any heating effects at
radii interior to $r_{\rm t,p}$ due to the time variation of the tidal
field. In contrast to a King model, the Plummer density law
(\ref{eqn:plummer}) formally extends to infinity and so the outer
radius of a cluster is not well-defined. The length unit we use yields
a pericentric tidal radius close to the (random) position of the
outermost star in our initial clusters. The internal stellar
velocities are scaled to ensure virial equilibrium at the start of the
simulation.

The tidal radius of a cluster on an elliptical orbit does not have a
unique value. It varies continuously both as the cluster moves along
its orbit and when the cluster loses mass through stellar evolution or
the escape of member stars. However, for clusters whose internal
relaxation times are longer than their orbital periods, tidal
stripping tends to truncate the star cluster at $r_{\rm
t,p}$~\citep{ol92}. These two time scales are of similar magnitude for
our simulated clusters so we expect two-body relaxation processes to
lead to some expansion of the cluster beyond $r_{\rm t,p}$. We take
account of this by defining escaping stars to be those stars outside
twice the apogalactic tidal radius $r_{\rm t,a}$. Stars whose orbits
carry them outside this radius are generally lost to the
cluster~\citep[e.g.][]{he01}. Stars are thus free to populate the
region between $r_{\rm t,p}$ and $2 r_{\rm t,a}$. 

The choice of length unit determines the physical density of the
cluster and hence the physical rate at which two-body relaxation
processes affect the evolution of the cluster. Our aim is to
investigate the influence of a time-varying external tidal field so it
is important to ensure that all internal evolutionary processes
proceed in an identical manner for each of our simulated clusters. We
therefore use the same length unit for the clusters on circular orbits
as we chose for those on elliptical orbits. This means that the
clusters on circular orbits also do not initially fill their tidal
radii and so necessarily allows expansion of these clusters. However,
since the circular orbits are placed at the time-averaged radius of
the elliptical orbits this expansion should on average be the same for
both sets of clusters. Any differences in the rates of expansion are
due to the time variation in the tidal field.

The effect of steady tidal fields on star cluster evolution has been
the subject of papers by a number of
authors~\citep[e.g.][]{gh97,ah98,baum01,zw02}. However, until recently
the problem of clusters moving in time-varying tidal fields had
received relatively little attention from \nbody\,
simulations.~\cite{baum99} uses the results of some simulations of
single-mass star clusters orbiting a point mass to determine the
appropriate choice of escape radius for simulations of mass loss. More
recently,~\cite{bama02} presented an extensive survey of the dynamical
evolution of star clusters in external tidal fields. Our investigation
is complementary to theirs because we are looking at the evolution of
the observable global structure of the cluster while~\cite{bama02}
focus their attention mostly on the evolution of the stellar mass
function within the cluster.  The results of all our simulations are
presented in terms of quantities which can be compared directly with
those calculated from the observed data, thereby enabling us to
discuss how different dynamical processes impact on observations.

We follow the evolution of the clusters for about $5\, t_{\rm rh}$
(where $t_{\rm rh}$ is the half-mass relaxation time) This corresponds
to $T_{\rm max} \approx 1\,$Gyr in most cases. By this time, the
clusters on elliptical orbits have lost more than $50$ per cent of
their stars making estimates of properties such as the core radius
increasingly noisy. At regular intervals during the evolution, the
data from the clusters are recorded and all relevant properties of the
clusters are calculated.

\subsection{Results}
\label{sec:tides_results}
\begin{figure}
\includegraphics[width=84mm]{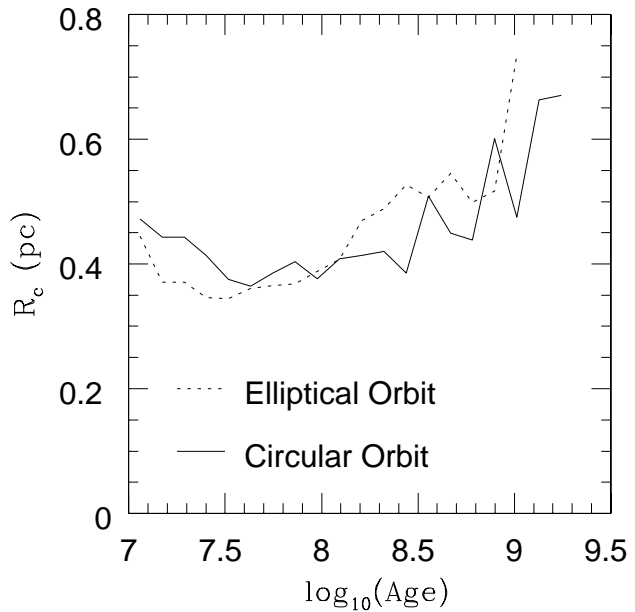} 
\caption{Core radius evolution for \nbody\, simulations of clusters on
orbits about a point mass representing the LMC. See text for
discussion.}  
\label{fig:circecc}
\end{figure}
\begin{figure}
\includegraphics[width=84mm]{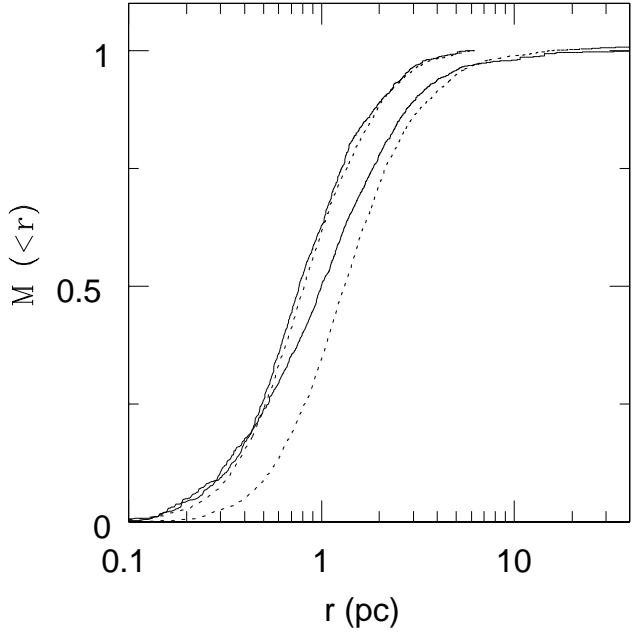} 
\caption{Cumulative mass profiles for stars above (solid curves) and
below (broken curves) the median stellar mass for a simulated cluster
containing only single stars and on a circular orbit about the
LMC. Profiles at two different times are shown: $T = 0$ (upper curves)
and $T=10.5\,$Myr (lower curves). There is clear evidence for mass
segregation developing in the cluster.}
\label{fig:massseg}
\end{figure}
The solid curve in Fig.~\ref{fig:circecc} presents the evolution of
the core radius for clusters on circular orbits about the LMC at a
distance of about $6\,$kpc from the centre. The broken curve shows the
evolution of $R_{\rm c}$ for clusters on elliptical orbits with the
same time-averaged galactocentric radius. It is apparent from this
figure that the core radius evolution in both cases is
indistinguishable, especially in view of the noise associated with the
output of small-$N$ simulations. During the first $10^8\,$yrs, the
core radius is seen to decrease slightly from its initial value. This
is because the calculation of $R_{\rm c}$ is dominated by the light of
the most massive stars which rapidly become more centrally
concentrated due to the effects of mass segregation. As
Fig.~\ref{fig:massseg} shows, after $10.5\,$Myr the half-mass radius
of the more massive stars is already $30$ per cent smaller than that
of the less massive stars.

\begin{figure}
\includegraphics[width=84mm]{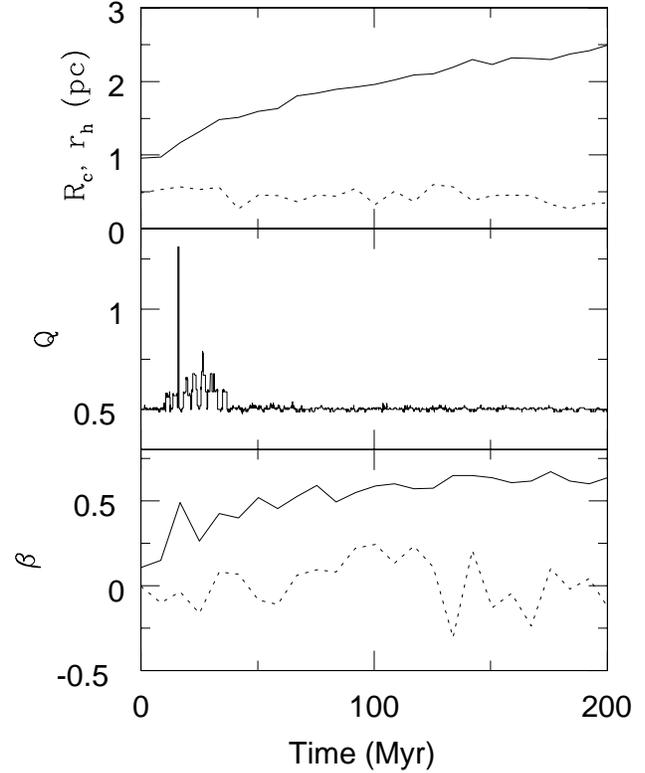} 
\caption{Early evolution of the velocity structure of a cluster on a
circular orbit. Top: evolution of the half-mass radius $r_{\rm h}$
(solid curve) and core radius $R_{\rm c}$ (broken curve); Middle: plot
of the virial ratio $Q$ (whose equilibrium value is $0.5$); Bottom:
velocity anisotropy parameter $\beta$ of stars within the inner $10$
per cent Lagrangian radius (broken curve) and between the $80$ per
cent and $90$ per cent radii (solid curve). See text for discussion.}
\label{fig:velevol}
\end{figure}

\begin{figure}
\includegraphics[width=82mm]{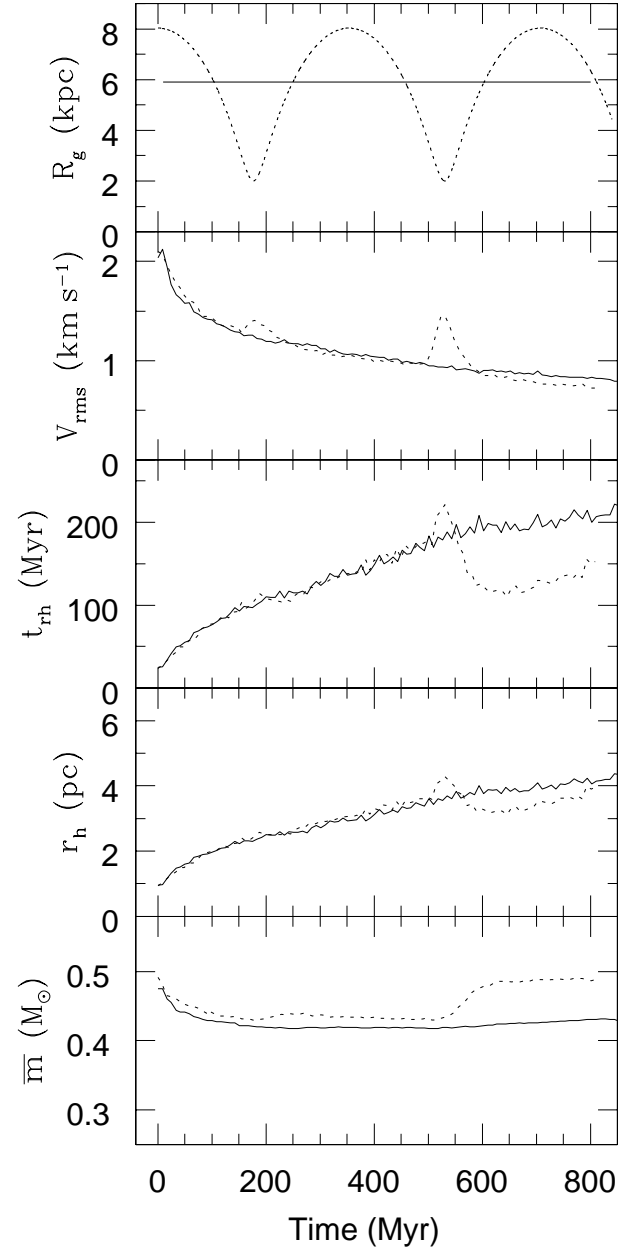} 
\caption{(Top panel) Variation of galactocentric radius $R_{\rm g}$
with time for a cluster on an elliptical orbit (broken curve). The
solid line shows the position of the circular orbit of equal
time-averaged radius. (Lower four panels) Evolution of rms velocity
$v_{\rm rms}$, half-mass relaxation time $t_{\rm rh}$, half-mass
radius $r_{\rm h}$ and mean stellar mass $\overline{m}$. Curves for
two cases are shown: (1) a cluster on a circular orbit (solid curves)
and (2) a cluster on an elliptical orbit (broken curves). Note that
the half-mass radius is calculated using all stars within a radius of
$2 r_{\rm t,a}$ of the cluster centre. In the simulations all such
stars are classified as non-escapers, although at any given time a
fraction of them are formally unbound.}
\label{fig:internal_props}
\end{figure}

After its initial contraction phase, the core radius begins to
increase. This occurs because the clusters do not fill the tidal
radius corresponding to their galactocentric positions at the start of
the simulations. A combination of mass loss from the rapid evolution
of massive stars and two-body relaxation can therefore drive the
overall expansion of the cluster seen in Fig.~\ref{fig:massseg}. The
top panel of Fig.~\ref{fig:velevol} shows that the cluster's half-mass
radius $r_{\rm h}$ is increasing with time (see fourth panel of
Fig.~\ref{fig:internal_props} for long-term $r_{\rm h}$ behaviour).
During this period its $r_{\rm d}$ (not shown) is roughly constant at
about $0.3\,$pc.

The expansion of $r_{\rm h}$ is driven by a combination of two
processes which are illustrated in the lower panels of
Fig.~\ref{fig:velevol}. The middle panel shows the evolution of the
virial ratio $Q$. Initially the cluster is in equilibrium with $Q =
0.5$. After about $10\,$Myr, the cluster is temporarily displaced from
equilibrium ($Q > 0.5$) by mass loss from the most massive stars. This
is accompanied by the increase in the half-mass radius of the cluster
shown in the top panel of Fig.~\ref{fig:velevol}. It is interesting to
note that the $R_{\rm c}$ of the cluster does not change as a result
of this mass loss. Stellar evolution cannot account for the sustained
expansion of the cluster, however. The cluster loses about $5$ per
cent of its mass as a result of the above mass loss events. Using the
virial theorem it is straightforward to show that the gravitational
radius $R^\prime$ of a stellar system with initial radius $R$ and mass
$M$ which loses an amount of mass $m$ is given by~\citep{hills80}
\begin{equation}
R^\prime = \frac{1}{2}\left(\frac{M-m}{\frac{1}{2}M - m}\right) R .
\end{equation}
A $5$ per cent mass loss should lead to an increase of less than $10$
per cent in the size of the cluster which is insufficient to account
for the cluster expansion seen in Figs.~\ref{fig:circecc}
and~\ref{fig:massseg}.

The bottom panel of Fig.~\ref{fig:velevol} shows the evolution of the
velocity anisotropy parameter $\beta = 1 - \langle v_{\rm t}^2 \rangle
/ 2 \langle v_{\rm r}^2\rangle$~\citep{bt87} inside the $10$ per cent
Lagrangian radius of the cluster (the radius containing $10$ per cent
of the total mass of the cluster) and in the outer parts (between the
$80$ per cent and $90$ per cent Lagrangian radii). Initially, the
velocity distribution throughout the cluster is isotropic ($\beta =
0$). While the central parts of the cluster retain this isotropy, in
the outer parts the distribution becomes increasingly radially
anisotropic ($\beta > 0$). This indicates that the continued growth of
$r_{\rm h}$ is being fueled by encounters in the inner parts of the
cluster which produce a population of stars on extended radial
orbits. The velocity distribution in the outer parts of isolated
clusters simulated by~\cite{bhh02} exhibited similar behaviour during
post core-collapse evolution; those authors similarly concluded that
the growth of the outer Lagrangian radii at late times was due to
two-body processes in the inner regions.

Given the similarity between the core radius evolution of clusters on
circular and elliptical orbits, we now proceed to investigate the
extent to which the clusters on elliptical orbits have been affected
by the variation in the tidal field. Fig.~\ref{fig:internal_props}
presents the evolution of certain key internal properties of two
typical clusters, one on a circular orbit and one on an elliptical
orbit. The top panel shows the galactocentric radius $R_{\rm g}$ along
the elliptical orbit to emphasize the clear correlation between
perigalactic passages and changes in the internal properties of the
cluster on the elliptical orbit. The second panel shows that, as
expected, during a perigalactic passage the rms velocity $v_{\rm rms}$
of the stars in the cluster increases sharply. This velocity kick
ejects the most energetic stars from the cluster. As a result of this
mass loss, $v_{\rm rms}$ settles to a value slightly below that of the
cluster on the circular orbit once it returns to virial equilibrium.

The third and fourth panels of Fig.~\ref{fig:internal_props} show how
$t_{\rm rh}$ and $r_{\rm h}$ evolve during the cluster orbit. The most
noticeable effect is that the relaxation time for the cluster on the
elliptical orbit is significantly reduced after the second
perigalactic passage, so that the cluster evolves more rapidly
following this encounter. This is due both to the decrease in the
total number of stars in the cluster and to the increase in the mean
stellar mass $\overline{m}$ (shown in the bottom panel of
Fig.~\ref{fig:internal_props}). The increase in $\overline{m}$ occurs
because, on average, low-mass stars acquire larger velocities during
pericentre passages and are hence more likely to escape from the
cluster. The value of $r_{\rm h}$ is seen to increase during the
perigalactic passage but rapidly settles back to a value slightly
below the circular-orbit value. Thus while the tidal shocks
experienced by the clusters on elliptical orbits do not affect the
evolution of $R_{\rm c}$, they are nonetheless strong enough to modify
other important properties of the cluster. As
Fig.~\ref{fig:internal_props} shows, the impact of the shocks on the
cluster properties increases with time -- this is due to the
decreasing number of bound stars in the cluster.

\subsection{Strength of tidal shocks}
\label{sec:shock_strength}
\begin{figure}
\includegraphics[width=84mm]{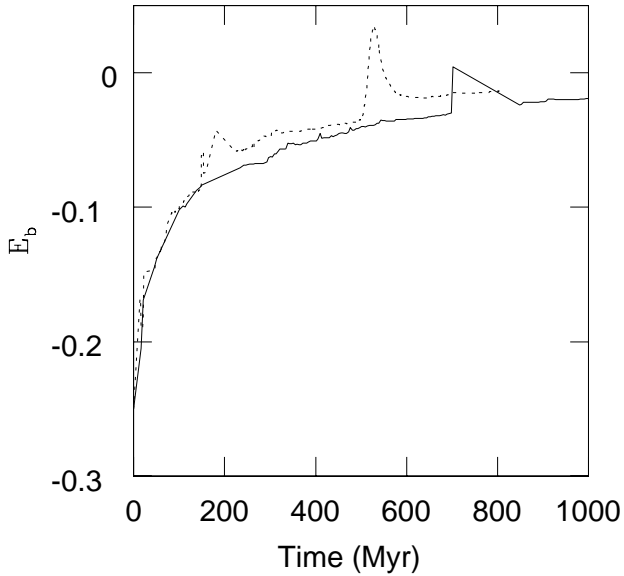} 
\caption{Evolution of the energy binding the cluster $E_{\rm
b}$. Curves for two cases are shown: (1) a cluster on a circular orbit
(solid curve) and (2) a cluster on an elliptical orbit (broken
curve). Energies are given in standard \nbody units ($G = M_0 = 1,
E_{{\rm b},0}= -1/4$).}
\label{fig:energy}
\end{figure}

Before drawing our final conclusions based on these
simulations, we now investigate whether the lack of impact on $R_{\rm
c}$ is due to the weakness of the shocks we have simulated or whether
the value of $R_{\rm c}$ is intrinsically robust to tidal
shocking. The strength of a tidal shock is generally presented in
terms of the dimensionless parameter $\beta_{\rm sh}$ given
by~\citep[e.g.][]{spit87}
\begin{equation}
\beta_{\rm sh} = \frac{t_{\rm rh}}{t_{\rm sh}},
\end{equation}
where $t_{\rm sh}$ is an estimate of the destruction time of the
cluster owing to energy input from the gravitational shocks. The time
scale $t_{\rm sh}$ is given by
\begin{equation}
t_{\rm sh} = \frac{\vert E_{\rm h} \vert}{d E_{\rm h}/dt},
\end{equation}
where $E_{\rm h}$ is the energy of the cluster evaluated at the
half-mass radius and $d E_{\rm h}/dt$ is the rate at which stars at
the half-mass radius acquire energy due to the shock. In a simulation
containing a finite number of stars, evaluation of this expression
requires some care. Averaging the energy changes of stars near the
half-mass radius does not produce robust results because individual
stars may gain and lose energy during the perigalactic passage through
other means than interaction with the external tidal field. Instead we
calculate the total energy binding the cluster as a function of time
and estimate the energy input to the cluster during perigalactic
passages. Fig.~\ref{fig:energy} compares the time evolution of the
binding energy $E_{\rm b}$ of a cluster on a circular orbit with that
of a cluster on an elliptical orbit. For a cluster in an external
potential $E_{\rm b}$ is defined as~\citep[e.g.][]{fh95}
\begin{equation}
E_{\rm b} = {\rm T} + {V} + {\rm E_{\rm T}}
\end{equation}
where $T$ and $V$ are the total kinetic and potential energy of the
cluster, respectively, and $E_{\rm T}$ is the energy in the tidal
field. The binding energy must be negative for the cluster to remain
bound. The effects of the external field are visible as dramatic
changes in $E_{\rm b}$ at about $180$ and $520\,$Myr. Note that the
sudden, temporary change in $E_{\rm b}$ for the circular run at about
$700\,$Myr is due to a collision between two binaries that results in
the the break-up of one of the binaries and the subsequent ejection of
its components. The ratio of the energy produced by this event to the
binding energy of the cluster emphasises the relatively fragile nature
of the clusters we are simulating.

\begin{figure}
\includegraphics[width=84mm]{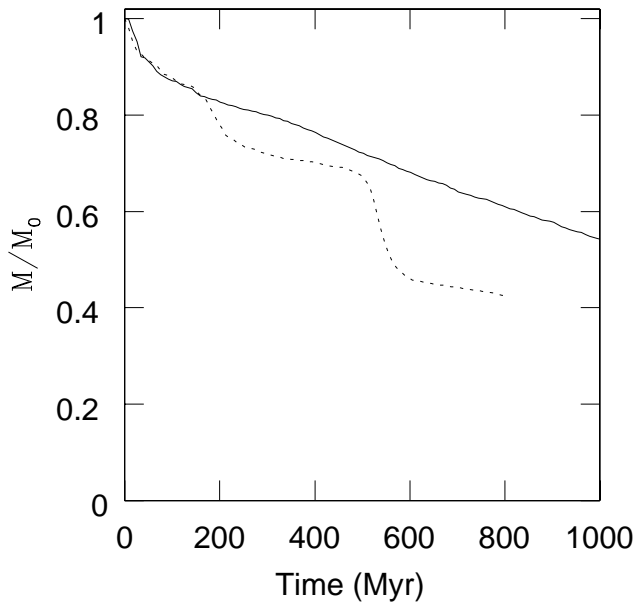} 
\caption{Evolution of total cluster mass $M$ for clusters on circular
(solid curve) and elliptical (broken curve) orbits in units of the
initial mass $M_0$. Enhanced mass loss during the period of
revirialisation which follows each perigalactic passage is seen in the
elliptical orbit case.}
\label{fig:masstime}
\end{figure}
During perigalactic passages, some stars acquire velocities above the
escape velocity of the cluster and this leads to an enhanced mass-loss
rate (see Fig.~\ref{fig:masstime}). However, the finite time required
for such stars to escape means that $E_{\rm b}$ can temporarily become
positive because it is calculated by summation over all stars within
twice the tidal radius of the cluster. After the cluster has passed
pericentre, stars continue to leave across the tidal boundary and
$E_{\rm b}$ returns to a negative value indicating that the remaining
cluster is still gravitationally bound. Following each shock, the
cluster undergoes a revirialisation process during which the energy
acquired from the tidal interaction is redistributed among the cluster
stars. This process results in further loss of mass from the cluster
because, on average, the revirialisation process gives velocity kicks
to stars which are three times larger than those arising from the
initial shock~\citep[see e.g.][]{go99}.

\begin{figure}
\includegraphics[width=84mm]{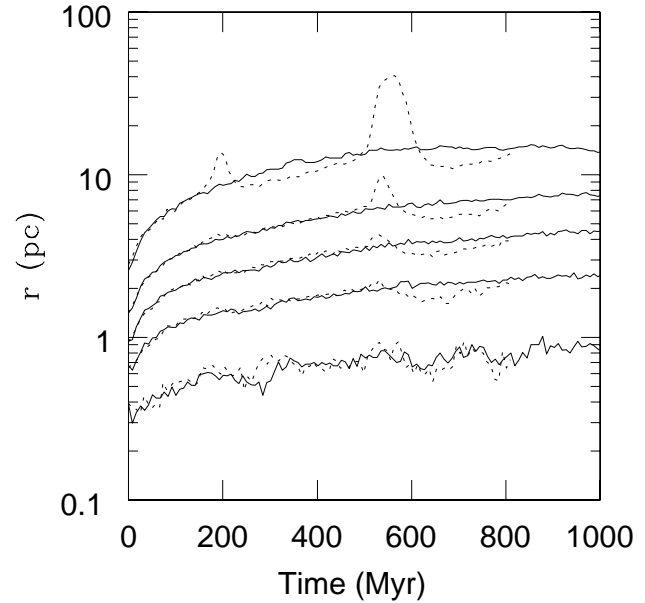} 
\caption{Evolution of the Lagrangian radii. Curves for two cases are
shown (1) for a cluster on a circular orbit (solid curves) and (2) for
a cluster on an elliptical orbit (broken curves). From bottom to top,
the curves show the radii containing $10$, $30$, $50$, $70$ and $90$
per cent of the cluster mass. The heating effect from the external
tidal field is evident, even at the radius containing only $30$ per
cent of the cluster mass.}
\label{fig:massradcomp}
\end{figure}
We take the energy input by the shock to be the difference between the
energy of the cluster after this revirialisation process is complete
and the energy of the cluster on the circular orbit at this time
(about $T = 540\,$Myr). The time between the shocks is $340\,$Myr,
which leads to $t_{\rm sh} \approx 6800\,$Myr. At this time $t_{\rm
rh}$ for the cluster is about $170\,$Myr, so the shock strength is
about $0.02$. Comparing this value with Fig.~13 of~\cite{glo99} and
noting that the initial concentrations $c = \log_{10}(r_{\rm t}/r_{\rm
d})$ of our clusters are about $1.1$ (where we assume that $r_{\rm t}$
is the initial outer limit of the clusters), we find that our clusters
lie approximately on the borderline of the region where shocks are
important. We conclude that the shocks our clusters are experiencing
would significantly affect their evolution on long time
scales. Indeed, as Fig.~\ref{fig:massradcomp} illustrates, even the
$30$ per cent Lagrangian radius of the cluster is affected by the
tidal shocks, indicating energy input into the cluster over a wide
range of radii. Nevertheless, on a timescale of up to $10^9$ yrs, the
effects of tidal heating are insufficient to modify the observed
$R_{\rm c}$ of the clusters significantly.

\subsection{Discussion}
\label{sec:tides_discussion}
\begin{figure}
\includegraphics[width=84mm]{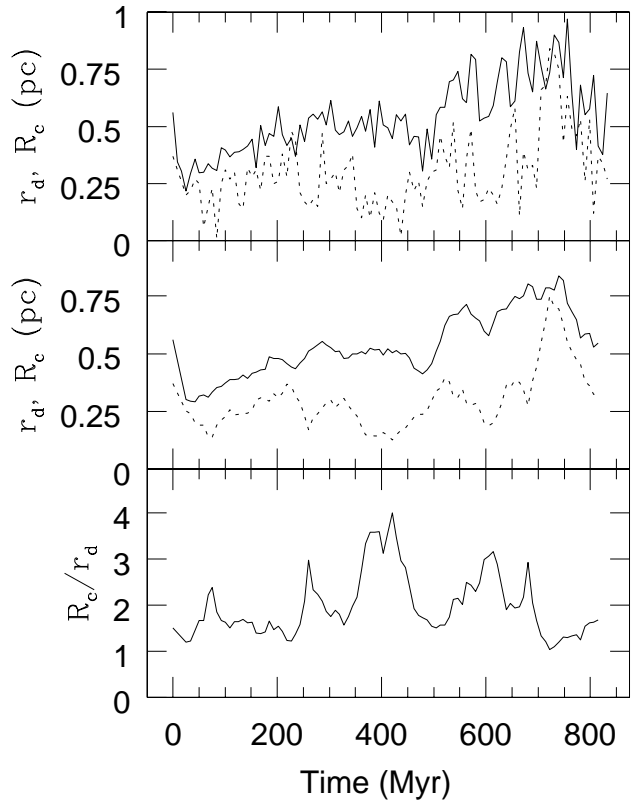} 
\caption{Comparison of the evolution of the density radius $r_{\rm d}$
(broken curve) with the observational core radius $R_{\rm c}$ (solid
curve) for a single \nbody cluster. Top: Un-smoothed results from the
simulations. Middle: Curves smoothed by averaging over five time
bins. Bottom: Ratio of $R_{\rm c}$ to $r_{\rm d}$ calculated using the
smoothed curves from the middle panel. See text for discussion.}
\label{fig:corecomp}
\end{figure}
The top panel of Fig.~\ref{fig:velevol} illustrates an important point
which is worth emphasizing in the context of comparisons between
simulated data and observations, namely that $R_{\rm c}$ and $r_{\rm
h}$ are not, in general, related in a simple way. In particular
circumstances, for example at the start of our simulations where the
model clusters are in strict dynamical equilibrium, it is easy to show
that, for a Plummer model composed of equal mass stars, the two
quantities are related by $r_{\rm h} = 2 R_{\rm c}$. However,
Fig.~\ref{fig:velevol} shows that this relation quickly breaks down as
the cluster evolves away from its initial conditions. Thus, in the
absence of further information, observational estimation of $R_{\rm
c}$ must not be assumed to imply knowledge of the mass distribution in
the cluster.

The same is true of the density radius $r_{\rm d}$ defined in
Section~\ref{sec:introduction}. Fig.~\ref{fig:corecomp} shows the
evolution of $R_{\rm c}$ (solid curve) and $r_{\rm d}$ (broken curve)
for one of the clusters on an elliptical orbit. As the top panel of
the figure shows, in a small-$N$ simulation, both $R_{\rm c}$ and
$r_{\rm d}$ are intrinsically noisy quantities: this is emphasised by
the middle panel where we have smoothed the curves from the top panel
using a window function of width five time bins. The bottom panel of
Fig.~\ref{fig:corecomp}, which shows the ratio of $R_{\rm c}$ to
$r_{\rm d}$ for the smoothed data, demonstrates that the relation
between the two quantities is not a simple proportionality and that
the evolutionary trends in the two quantities can be quite
different. The complicated relation between light and mass in star
clusters is the result of a variety of processes such as stellar
evolution and mass segregation which modify the mass and light
distributions in different ways. Given the absence of simple relations
between the different core radii, great care must be exercised when we
interpret observational data through comparison with numerical
simulations or theoretical models.

The simulations we have performed contain relatively low numbers of
stars and are smaller and of lower total mass than the actual LMC
clusters. However, the mean densities of our clusters are very similar
to those of the LMC clusters and it is this which determines the
extent to which an external tidal field can influence the internal
dynamics of a cluster. On this basis, we can be confident that
conclusions based on our simulations are relevant to the clusters in
the LMC. Our simulations have shown that, on a timescale of $1\,$Gyr,
which corresponds to several $t_{\rm rh}$ for our model clusters,
tidal forces are not able to produce a spread in core radius similar
to that observed among the LMC clusters.

There are two obvious avenues for building on the simulations we have
performed. First, simulations of clusters containing a larger numbers
of stars $N$ could be performed. This would allow us to achieve
cluster masses comparable to those of the real LMC clusters. However,
increasing $N$ in a cluster while holding the physical density of the
cluster constant leads to an increase in $t_{\rm rh}$. The expansion
seen in $R_{\rm c}$ for our simulated clusters is driven primarily by
two-body processes (which proceed on a time scale $t_{\rm rh}$) rather
than by tidal heating and it will therefore be slowed by increasing
$N$. Further, the disruptive effect of pericentre passages depends on
the ability of a cluster to re-fill the region outside its pericentric
tidal radius $r_{\rm t,p}$ during the time between close passages. As
$t_{\rm rh}$ increases this re-filling is reduced. The cluster will
therefore have fewer stars outside $r_{\rm t,p}$ and hence will be
less affected by shocks than small-$N$ clusters (see Fig.~2
of~\cite{bama02}). A cluster on an elliptical orbit and which, as in
our simulations, initially filled its pericentric tidal radius would
remain truncated at this radius throughout its evolution because there
would not be sufficient time between perigalactic passages for the
region outside $r_{\rm t,p}$ to be re-filled~\citep{ol92}. Thus shocks
should have less impact on the $R_{\rm c}$ of larger-$N$ clusters and
the observed expansions of clusters on both circular and elliptical
orbits should remain similar.

The second possible improvement would be to use a more realistic model
of the LMC potential, for example one based upon the recent mass
models of~\cite{vdm02}. However, the actual LMC mass distribution is
more extended than the Keplerian potential we have been using in our
simulations and therefore the gradients in its gravitational field are
smaller. Consequently, the associated strength of its tidal effects on
clusters moving on elliptical orbits is reduced.

Thus, we conclude that neither of the above modifications to our
simulations would tend to increase the effect of the tidal field. The
fact that the time-varying tidal field in our simulations did not
result in changes in the evolution of $R_{\rm c}$ in the extreme case
of a low-mass cluster in orbit about a point-mass galaxy leads us to
conclude that tides have not yet had time to play a significant role
in the internal evolution of the intermediate-age LMC clusters.

\section{Models with primordial binaries}
\subsection{Steady tidal field case}
\label{sec:binfrac}
In this section we present results for models moving in a steady tidal
field but with different numbers of primordial binaries, in order to
isolate the impact of large binary fractions on the evolution of the
cluster properties. The clusters discussed here move on the same
orbits as the circular orbit runs of Section~\ref{sec:tides}.

Based on an analysis of the observed binary sequence in the young
cluster NGC 1818,~\cite{el98} have shown that the binary fraction
$f_{\rm b}$ in the centre of this cluster is about $0.35$. Here
$f_{\rm b} = N_{\rm b}/(N_{\rm s}+N_{\rm b})$,where $N_{\rm s}$ and
$N_{\rm b}$ are the numbers of single stars and binary systems,
respectively. To date, the binary fractions of the older clusters in
our sample have not been determined. Thus, at present the binary
fractions of these clusters are essentially free
parameters. Heggie's~\citeyearpar{heg75} law implies that only hard
binaries are dynamically important in the evolution of a star
cluster. A binary is said to be hard when the magnitude of its binding
energy $\vert E_{\rm bin}\vert$ exceeds the kinetic energy of a
typical cluster star~\citep{heg75}. In the case of a binary with
component masses $m_1$ and $m_2$ at a separation $a$ moving in a
cluster with mean mass $\overline{m}$ and velocity dispersion $\sigma$
the hardness condition is
\begin{equation}
\frac{G m_1 m_2}{2 a} > \frac{1}{2}\overline{m} \sigma^2.
\end{equation}
(We note that the NBODY4 code uses a dispersion of $\sqrt{2}\sigma$ in
this equation to determine the hardness limit for binaries, in order
to take account of the larger velocity dispersion in the core of a
centrally concentrated cluster.) In our simulations the boundary
between hard and soft binaries occurs at a separation of about
$120\,$AU for binaries containing stars of the mean stellar mass in
the cluster. Hard binaries act as heat sources because they become
more tightly bound as a result of interactions with single stars and
thereby impart kinetic energy to the stars they encounter.~\cite{ha92}
found that the density radii of clusters containing primordial
binaries experienced shallower contraction during core collapse and
subsequently tended to be larger than those of clusters containing
only single stars. This result, based on simulations of single-mass
clusters with binary fractions of about $0.03$, motivates our
investigation of clusters containing large numbers of primordial
binaries. Hereinafter, we will use $f_{\rm b}$ to refer to the
fraction of \emph{hard} binaries in the cluster.

As an extreme case, we consider first clusters with initial $f_{\rm b}
= 0.5$ and with the hard binaries randomly distributed throughout the
cluster. In each binary the mass of the primary star is drawn from the
IMF of~\cite{ktg93} and the mass of its companion is obtained by
randomly choosing a mass ratio for the binary from a uniform
distribution constrained by the single star mass limits~\citep[for a
more detailed discussion, see][]{htap01}. The binary orbital
parameters are generated from the distribution of~\cite{eft89}. The
clusters are otherwise generated exactly as in the single star runs of
Section~\ref{sec:tides}. We note that the inclusion of a large number
of hard primordial binaries dramatically increases the computation
time required for each simulation.

\begin{figure}
\includegraphics[width=84mm]{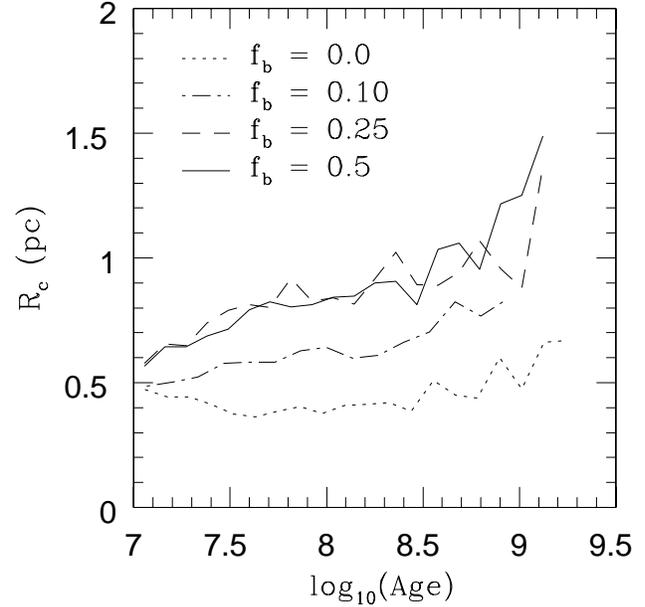} 
\caption{Core radius evolution for \nbody\, clusters containing single
stars only (short-dashed curve) and clusters containing $10$ per cent,
$25$ per cent and $50$ per cent hard primordial binaries (short-long
dashed, long-dashed and solid curves, respectively). All clusters are
on circular orbits about the LMC. See text for discussion.}
\label{fig:sinbin}
\end{figure}
\begin{figure}
\includegraphics[width=84mm]{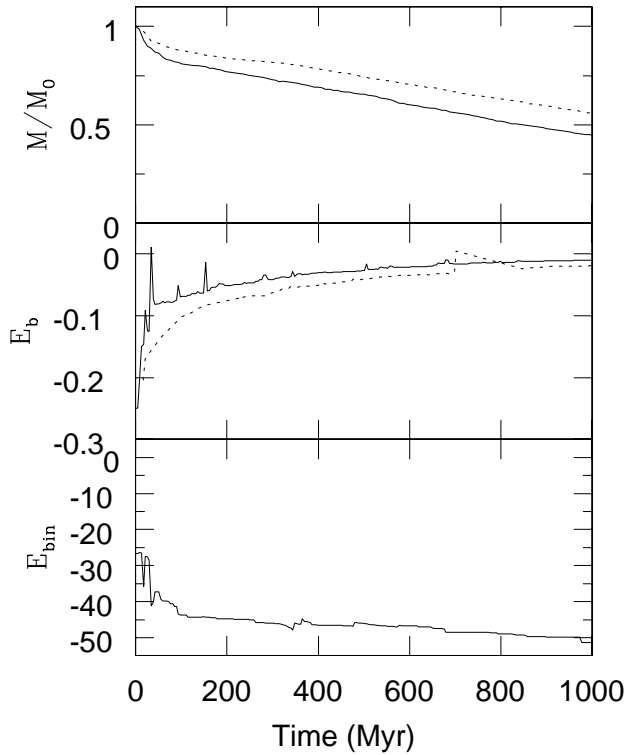} 
\caption{Comparison of the evolution of the total cluster mass $M$
(top panel) and the energy binding the cluster $E_{\rm b}$ (middle
panel). Two cases are shown (1) a cluster without primordial binaries
(broken curves) and (2) a cluster containing $50$ per cent primordial
binaries (solid curves). The total mass is normalised by its initial
value $M_0$ and $E_{\rm b}$ is given in standard \nbody units. The
bottom panel shows the evolution of the internal binding energy of the
primordial binaries $E_{\rm bin}$ (in standard \nbody units) in the
run containing primordial binaries. The decrease in $E_{\rm bin}$ is
the result of interaction-induced binary hardening. Note that $E_{\rm
bin}$ includes the internal energy of escaped binaries. See text for
discussion.}
\label{fig:sinbincomp}
\end{figure}
\begin{figure}
\includegraphics[width=84mm]{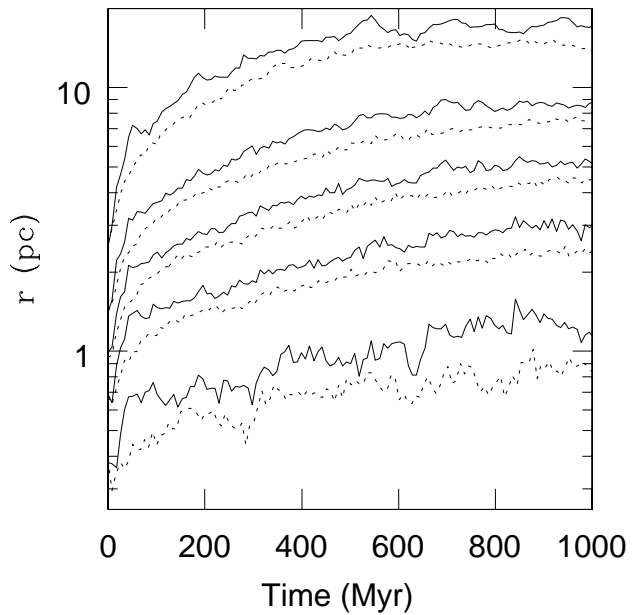} 
\caption{Evolution of the Lagrangian radii for clusters with $50$ per
cent primordial binaries (solid curves) and with single stars only
(broken curves). From bottom to top, the curves show the radii
containing $10$, $30$, $50$, $70$ and $90$ per cent of the cluster
mass. The heating effect of the primordial binaries is visible at all
radii.}
\label{fig:sinbinpercent}
\end{figure}
Fig.~\ref{fig:sinbin} compares the evolution of $R_{\rm c}$ for
clusters containing $N=5000$ single stars and no primordial binaries
with that for clusters containing a significant initial binary
population. An increased binary fraction produces a noticeable effect
on the evolution of $R_{\rm c}$. After $1\,$Gyr, the core radii of
clusters containing $50$ per cent primordial binaries are about a
factor of two larger than those of clusters with no binaries. 

The results presented in Fig.~\ref{fig:sinbincomp} give some insight
into the mechanism by which a large binary fraction can drive an
expansion of $R_{\rm c}$. The top panel shows that the rate at which
mass is lost from a cluster with initial $f_{\rm b} = 0.5$ is
significantly greater than in the case of a cluster with no initial
binary population. After $1\,$Gyr this cluster has lost almost $25$
per cent more mass than the cluster containing only single stars. The
evolution of the energy binding the cluster $E_{\rm b}$ is shown in
the second panel of Fig.~\ref{fig:sinbincomp}. In a cluster with a
large initial binary fraction, $E_{\rm b}$ rapidly attains a value
significantly higher than that of a cluster of single stars. The
enhanced mass-loss rate due to the presence of the binaries decreases
the depth of the cluster's potential-well; the magnitude of the
binding energy is further reduced by the heating of the remaining
bound stars through weak encounters. As Fig.~\ref{fig:sinbinpercent}
shows, the energy input due to binary heating is distributed
throughout the cluster. All the Lagrangian radii of the cluster
rapidly attain larger values when the cluster contains primordial
binaries. We note that, as was discussed in
Section~\ref{sec:tides_results}, the growth of the clusters containing
only single stars is due to the fact that the clusters do not fill
their tidal radii at the start of the simulations. We use identical
initial conditions for the clusters containing binaries -- the excess
heating observed in those clusters is due to the presence of the hard
binary population.

The third panel of Fig.~\ref{fig:sinbincomp} shows the internal
binding energy of the primordial hard binaries $E_{\rm bin}$ as a
function of time. In calculating $E_{\rm bin}$ we include the binding
energy of those hard binaries which have escaped from the cluster
because we are interested in the rate at which encounters lead to the
hardening of binaries. This plot demonstrates that, as expected, the
hard binaries become harder with time and so impart energy to the
cluster. It is well known that energy generation by hard binaries
halts core collapse~\citep[e.g.][]{ha92}. However,
Fig.~\ref{fig:sinbincomp} demonstrates that even in the very early
stages of a cluster's evolution, the heating effect of a large
primordial hard binary population can have a significant impact on the
energy budget of the cluster. 

We note that the occasional large changes in $E_{\rm bin}$ are the
result of stellar evolutionary processes and therefore do not
contribute to the direct heating of the cluster. For example, at
$T=34\,$Myr the components of one binary system merge to form a single
star: prior to the merger the binary is hardened by tidal interactions
between its components. The massive star thus produced explodes as a
supernova soon after and the velocity kick received by the remnant as
a result of the explosion is reflected in the spike in $E_{\rm
b}$. The prominence of this event is due to the relatively small
number of stars in the simulation.

Fig.~\ref{fig:sinbin} shows that the $R_{\rm c}$ evolution of clusters
with $f_{\rm b} = 0.25$ is very similar to that of clusters with
$f_{\rm b} = 0.5$ initially. This suggests that the heating effect of
primordial binaries saturates above a certain value of the binary
fraction $f_{\rm b,crit}$. For larger values of $f_{\rm b}$, the
$R_{\rm c}$ evolution is unchanged, although the mass-loss rate is
greater for increased $f_{\rm b}$. From Fig.~\ref{fig:sinbin} we see
that $f_{\rm b,crit}$ for the clusters we have simulated appears to
lie in the range $0.1-0.25$ because an initial binary fraction of $10$
per cent has a noticeably smaller impact on $R_{\rm c}$ than one of
$25$ per cent. We are currently investigating this effect in more
detail, including its dependence on the number of stars in the
simulation: the results will be presented elsewhere. Here we note only
that the saturation may arise from the fact that at early times only a
relatively small number of binaries take part in the heating
process. In our simulations only about three per cent of the binaries
are dynamically hardened during the first $100\,$Myr. We find that
this fraction appears to be independent of the value of $f_{\rm b}$.
Whether a particular binary experiences an encounter which leads to
hardening depends on the masses and separation of the components as
well as the local stellar density in the neighbourhood of the
binary. For a fixed binary mass, however, the hardest binaries have
the smallest separations and hence are least likely to interact with
other stars. This is reflected in our simulations: for a given binary
mass, all the binaries which undergo early hardening are initially
among the more weakly bound pairs. 

\subsection{Time-varying tidal field case}
\label{sec:bintide}
\begin{figure}
\includegraphics[width=84mm]{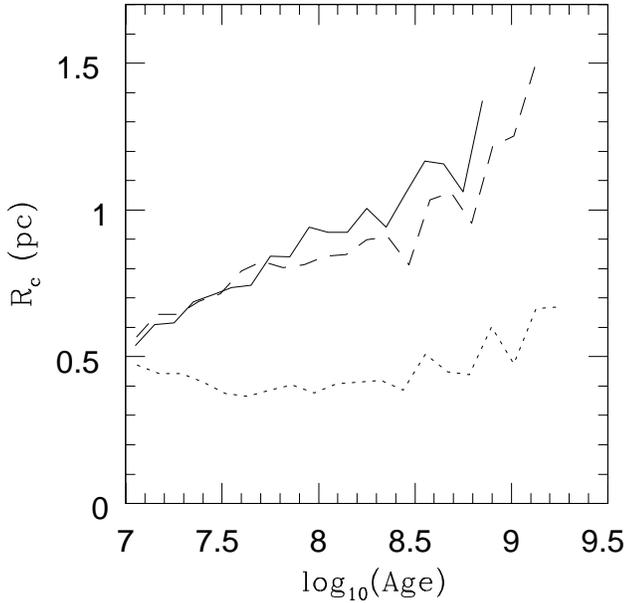} 
\caption{Core radius evolution for \nbody simulations of clusters on
different orbits about a point mass representing the LMC with (solid
and long-dashed curves) and without (short-dashed curve) hard
primordial binaries. The solid curve represents a cluster on an
elliptical orbit, while both broken curves represent clusters on
circular orbits. See text for discussion.}
\label{fig:eccbincorecomp}
\end{figure}
In our final set of simulations we examine clusters containing large
numbers of primordial binaries and moving on non-circular orbits about
the LMC. We have considered the extreme case of $f_{\rm b} = 0.5$ but,
as was shown in the previous section, these results also apply to
somewhat smaller values of $f_{\rm b}$. As
Fig.~\ref{fig:eccbincorecomp} shows, the core radius of a cluster on
an elliptical orbit and containing primordial binaries exhibits
stronger growth than that of a cluster of single stars on a circular
orbit: after about $600\,$Myr the core radius of the former cluster is
almost a factor of two larger. The increase in core radius is solely
due to the presence of binaries. As in the case of clusters containing
only single stars (Section~\ref{sec:tides}), a cluster with the same
primordial binary fraction but on a circular orbit exhibits almost
identical core radius evolution. However, the increased mass-loss rate
which results from binary heating leaves the cluster on the elliptical
orbit more vulnerable to tidal disruption. After only $600\,$Myr, the
cluster has lost almost $75$ per cent of its mass and the core radius
determination is consequently very uncertain. By $1\,$Gyr it is
virtually disrupted and retains only about $10$ per cent of its
initial mass while the circular orbit cluster still contains about
$45$ per cent of its initial mass.

\subsection{Discussion}
\label{sec:bintidedisc}
Our simulations demonstrate that the presence of a large population of
hard binaries can produce significant expansion of the core radius of
a cluster. In agreement with many previous
studies~\citep[e.g.][]{sm80,ha92,dlfm96} we have thus shown that
binaries constitute a dynamically significant population in star
clusters. In particular, we have found that primordial binaries can
rapidly affect observable cluster properties such as $R_{\rm c}$. In
the context of the LMC clusters, however, it is clear that the
presence of binaries alone cannot account for the the observed trends
in $R_{\rm c}$ and in particular for the factor of three difference in
core radius between NGC 1831 and NGC 1868. Even in the extreme case in
which one cluster has no primordial binaries at all while the other
cluster has a hard binary fraction of $25-50$ per cent, the difference
in $R_{\rm c}$ over the lifetime of the intermediate-age clusters is
at most a factor of two. The core radius of a cluster containing $10$
per cent binaries expands by only a factor of $1.5$ over the same time
scale. Further, the similarity of the observed present-day luminosity
functions of the clusters in the GO7307 sample~\citep{deg02a} argues
against large inter-cluster variations in the IMF and hence makes
significant differences in binary fraction unlikely.

The initial radial distribution of hard binaries has an impact on
their heating effect on the cluster stars~\citep[e.g.][]{dlfm96} and
may also have a bearing on the saturation of the heating effect which
we observed in Section~\ref{sec:binfrac}. A binary population which
was more centrally concentrated than the single stars as a result of
either primordial or dynamical mass segregation would be expected to
produce more heating because of the increased interaction rates in the
inner parts of the cluster. In our future simulations we will consider
alternatives to the uniform distribution used in the present
runs. However, based on our current simulations it is possible to
estimate the significance of centrally concentrating the binary
population. The clusters with $50$ per cent binary fractions contain
approximately the same number of binaries within $R_c$ as a cluster
with a $10$ per cent overall binary fraction in which all the binaries
are located within $R_c$. Given that none of our simulations exhibited
$R_c$ growth of a magnitude sufficient to match the observations, a
core binary fraction in excess of $50$ per cent would be required to
produce greater $R_c$ growth. The observed similarity of the degree of
mass segregation in the intermediate-age clusters~\citep{deg02a}
render it unlikely that the radial distribution of binaries could vary
so greatly between clusters.

As in the case of the simulations in Section~\ref{sec:tides}, it is
natural to ask whether our conclusions are also valid for clusters
containing larger numbers of stars than those in our current
simulations. The simulation of large-$N$ clusters containing
significant numbers of primordial binaries has only recently become
feasible and we are currently investigating the impact of binaries on
the evolution of such clusters; the results of this study will be
presented elsewhere. The present simulations show that the disruption
time for a small-$N$ cluster in a time-varying external tidal field is
significantly reduced if it contains a large fraction of primordial
binaries. It is not trivial to extrapolate these results to larger-$N$
clusters. However, binary heating of the cluster is essentially a
two-body process and therefore proceeds on the relaxation time scale
$t_{\rm rh}$. Given that a larger-$N$ cluster with the same mean
physical density has a larger $t_{\rm rh}$, the physical rate at which
the cluster is heated by its binary population will be reduced. Thus,
while larger-$N$ clusters would obviously survive for longer than
those in our present simulations, their physical rate of $R_{\rm c}$
expansion would be reduced. Large variations in primordial binary
fraction are therefore unlikely to be the explanation of the LMC
observations.

\section{Conclusions}
\label{sec:conclusions}
\begin{figure}
\includegraphics[width=84mm]{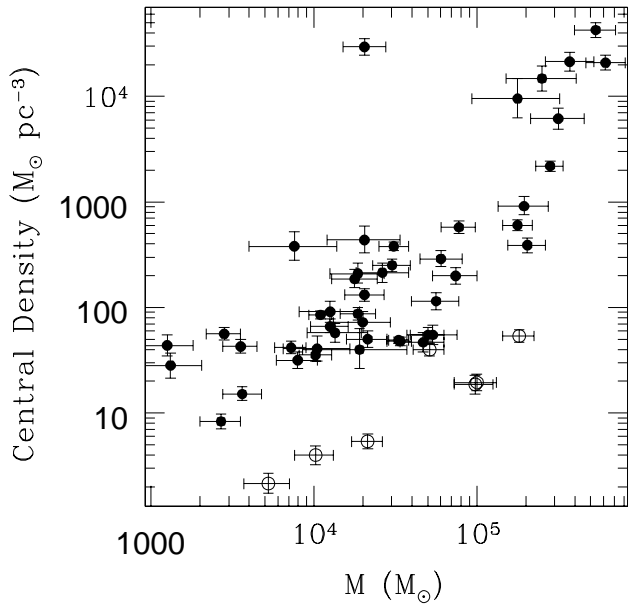}
\caption{Plot of central mass density versus estimated mass $M$ for
the LMC cluster sample. Clusters with $R_{\rm c}$ larger (smaller)
than $4$ pc are shown as open (filled) circles. Data are taken from
\protect\cite{mg03}.}
\label{fig:rhom}
\end{figure}
The core radii of star clusters in the LMC display a strong trend with
age, namely that the older clusters exhibit a much greater spread in
core radius than their younger counterparts. In particular, the
$R_{\rm c}$ of the intermediate-age clusters NGC 1868 and NGC 1831
differ by about a factor of three. It has been
suggested~\citep{efl89,mg03} that the trend represents a physical
evolution of the clusters and that the clusters all formed with
similar, small core radii which were modified by subsequent dynamical
evolution. The possibility that this might be caused by variations in
the cluster IMF has been ruled out based on HST photometry of a sample
of eight clusters~\citep{deg02a}. In this paper we have used \nbody\,
simulations of star clusters initially containing up to $N=5000$ stars
to investigate some possible alternative explanations. We have
considered two possible dynamical effects which could produce
significant $R_{\rm c}$ evolution: time-varying external tidal fields
and a varying population of primordial binaries.

Throughout this paper we have analysed our simulations using the same
procedures as were applied by~\cite{mg03} to the observed data. We
emphasised the importance of such an approach by comparing the
evolution of the observational core radius to that of the half-mass
radius and the density radius. As has been pointed out by a number of
authors~\citep[e.g.][]{zmhm01}, in general there are no simple
relations between these quantities. Thus observations of simulated
data sets are essential to ensure that comparisons between observed
and simulated data are made in a consistent manner.

We considered clusters on circular and elliptical orbits of equal
time-averaged radius about a point mass representing the LMC. The
clusters are scaled so that their limiting radii correspond to the
pericentric tidal radius of the clusters on elliptical
orbits. However, at the start of each simulation the clusters lie well
within the tidal radius which corresponds to their initial
galactocentric position (apocentre for the clusters on elliptical
orbits and the circular radius for the clusters on circular
orbits). Initially, the clusters expand due to stellar mass loss but
the competing effect of mass segregation means that $R_{\rm c}$
decreases slightly. At later times, two-body relaxation processes
drive a further expansion of the clusters because the clusters are not
tidally limited. This expansion is mirrored by a growth in $R_{\rm
c}$. However, the rate of $R_{\rm c}$ expansion is found to be
identical irrespective of the orbit on which the cluster is moving. We
conclude that differences in the cluster orbits cannot be responsible
for the observed differences in $R_{\rm c}$ among the LMC clusters.

Given that the binary fractions of all but the youngest clusters are
currently unconstrained by observations, we compared the evolution of
the $R_{\rm c}$ of a cluster containing no primordial binaries with
that of a cluster with a large initial hard binary fraction. The
presence of significant numbers of binaries leads to increased growth
of $R_{\rm c}$ compared with the evolution of clusters containing only
single stars. Binary fractions of $50$ per cent and $25$ per cent
produce similar degrees of $R_{\rm c}$ expansion, while a $10$ per
cent binary fraction results in noticeably less growth. Even in the
most extreme case of a cluster with a binary fraction of $50$ per cent
on an elliptical orbit, however, $R_{\rm c}$ increases by at most a
factor of two. Not only is this smaller than the increase required to
explain the LMC observations but the required spread in primordial binary
fraction is already difficult to reconcile with the observed
similarity of the cluster stellar luminosity functions~\citep{deg02a}.

Our simulations have demonstrated that neither large variations in
primordial binary fraction nor differences in cluster orbits about the
LMC can account for the $R_{\rm c}$--age relation seen in the LMC
clusters. The origin of the spread in $R_{\rm c}$ among the
intermediate-age clusters therefore remains an open issue. We now
return briefly to the observed data on the LMC cluster system as a
whole to see whether there are any further clues to the origin of the
trend.

Fig.~\ref{fig:rhom} presents evidence for a systematic difference in
the central density of the large-$R_{\rm c}$ clusters relative to the
rest of the cluster population: these clusters have systematically
lower central densities for a given cluster mass. This could be
indicative of variations in star formation efficiency (SFE) between
the clusters: low SFE leads to rapid cluster expansion and
consequently a drop in the central
density~\citep[e.g.][]{go97}. Simulations of cluster expansion due to
residual gas expulsion from clusters with low SFE~\citep{go97} do not
appear to show sustained core expansion of the magnitude seen in the
intermediate-age LMC clusters. However, the core radii quoted
in~\cite{go97} are those of the mass distribution in the simulated
clusters which, as was discussed in Section~\ref{sec:tides}, may not
accurately reflect the changes in the light distribution. We are
currently investigating the long-term evolution of clusters which
experience different degrees of primordial gas loss owing to SFEs in
the range $0.25-0.60$, required to explain the range of core radii
seen in the youngest LMC clusters~\citep{go97}. We are also simulating
the effects of primordial binaries on larger-$N$ clusters. Of
particular interest is the possibility that the heating effects of a
large binary population could enable a cluster to retain a large
initial $R_{\rm c}$ produced by gas expulsion.

\section*{Acknowledgments}
MIW acknowledges financial support from PPARC. JRH acknowledges
financial support from a Kalbfleisch Fellowship and thanks the
Institute of Astronomy (Cambridge, UK) for supporting a visit during
this work. ADM would like to acknowledge the support of a Trinity
College ERS grant and a British government ORS award. CAT thanks
Churchill College for a fellowship. We are very grateful to Sverre
Aarseth for assistance with the NBODY4 code and for numerous useful
discussions. We thank Pete Bunclark at the Institute of Astronomy,
Cambridge, for handling the installation and setup of the GRAPE-6
hardware. We also thank HongSheng Zhao for helpful discussions and
Richard de Grijs and Douglas Heggie for useful discussions and
comments on an earlier version of this paper.


\label{lastpage}

\end{document}